# Spectrally indistinguishable intermodal-vectorial four-wave mixing in birefringent few-mode fibers for spatial-polarization-frequency hybrid-entangled photon-pairs generation

**ANDRZEJ GAWLIK,* MARTA BERNAŚ, KINGA ŻOŁNACZ, AND KAROL TARNOWSKI**

Department of Optics and Photonics, Faculty of Fundamental Problems of Technology, Wrocław University of Science and Technology, 50-370 Wrocław, Poland

*andrzej.gawlik@pwr.edu.pl

**Abstract:**

In this paper, we use a birefringent few-mode fiber to demonstrate an intermodal-vectorial four-wave mixing process that generates two pairs of spectrally overlapping signal-idler bands. Using phase-matching conditions, we show that the pairs of bands become spectrally indistinguishable when the group refractive indices of the signal and idler modes intersect at the pump wavelength. Our theoretical predictions are confirmed through experimental observations corroborated by numerical simulations. Furthermore, we introduce the two-photon state associated with the overlapping bands, which exhibits hybrid entanglement in spatial-polarization-frequency degrees of freedom. We explain that its degree of entanglement can be manipulated by tailoring the spectral position of the overlapping pairs of bands with respect to the Raman band and varying the excitation ratio of the pump modes. This study introduces a scheme for a fiber-based source of photon pairs with spatial-polarization-frequency entanglement.

## 1. Introduction

The generation of entangled photon pairs plays a pivotal role in advancing modern quantum technologies, including quantum information processing and communication [1,2], quantum computing[3], and quantum cryptography[4,5]. Due to their compatibility with the existing communication networks, optical fibers represent a promising platform for the generation of entangled photon pairs through a nonlinear optical process of four-wave mixing (FWM). In FWM, the annihilation of two pump photons leads to the simultaneous creation of two (signal and idler) photons, which are entangled in the energy-time domain[6]. Furthermore, the signal and idler photons can exhibit additional quantum relationships, such as (i) correlations, (ii) hyper-entanglement, or (iii) hybrid entanglement, in one or more degrees of freedom (DOFs), including polarization, spatial distribution, or angular momentum[7]. The nature of these quantum relationships and the participating DOFs are determined by the modal structure of the fiber under consideration and the specific type of FWM process involved in the generation of the signal and idler photons. In essence, correlations are a natural consequence of the phase matching during FWM (e.g., the signal/idler are generated in specific polarizations or transverse modes), while entanglement of these correlated photons can be achieved through quantum interference, either in a Sagnac loop[8,9] or by spectrally overlapping distinct FWM processes[10,11]. Importantly, generating entanglement across multiple DOFs can increase the capacity of quantum communication and improve transmission across noisy channels.

The last decade has marked a tremendous progress in generating photon pairs that are either correlated or entangled across multiple DOFs using FWM. Different fiber designs have facilitated the theoretical predictions and experimental demonstrations of schemes utilizing various FWM processes to achieve such correlations and entanglement. In single mode fibers (SMFs), the vectorial FWM was used to demonstrate entanglement in frequency[8], polarization[9], as well as hyperentanglement in energy/time-frequency and energy/time-polarization[12]. As compared to SMFs, few-mode fibers (FMFs) offer the ability to guide higher-order modes thus extending the correlation/entanglement dimensionality by the spatial mode (also called transverse mode) degree of freedom. Different research groups used intermodal or intermodal-vectorial FWM in FMFs to demonstrate: spatial correlations[13,14] and entanglement[15], spatial-frequency correlations[10,16] and a scheme for such hybrid entanglement[10,11]. In our recent work[17], we demonstrated a new type of intermodal-vectorial four-wave mixing (IV-FWM), where the generated photons are correlated in the spatial-polarization-frequency DOFs[7].

In this manuscript, we propose a scheme for the generation of photon pairs with hybrid entanglement in the spatial-polarization-frequency DOF using a birefringent FMF. The entanglement is achieved by spectrally overlapping two pairs of signal-idler bands generated in a single intermodal-vectorial FWM process involving four distinct modes. This paper is organized as follows. In Section 2, we use the phase-matching conditions to derive the criteria for the existence of such spectrally overlapping two pairs of signal-idler bands and explain how their spectral positions can be tailored among others to avoid overlap with the Raman band. In Section 3, we provide an analytical expression for the gain of FWM and use it to discuss why the gain is primarily dependent on the overlap between the participating fields and not the spectral positions of the signal-idler bands. Subsequently, Section 4 presents the experimental demonstration of the aforementioned four-mode IV-FWM, which leads to the appearance of two pairs of spectrally overlapping signal-idler bands. These bands are accompanied by bands generated in two-mode IV-FWM process, which is correlated with them in the spatial-polarization-frequency DOFs. The experimental observations are corroborated by numerical simulations confirming the spatial and polarization nature of the participating modes. In Section 5, we define the two-photon state associated with the observed IV-FWM and explain how it can be engineered into a maximally entangled Bell state. Finally, Section 6 concludes the paper.

## 2. Phase-matching conditions

The process of four-wave mixing involves the conversion of two pump photons into idler and signal photons according to the scheme $(l, m) \rightarrow (p, n)$, where $l$ and $m$ refer to the pump modes, while $p$ and $n$ correspond to the signal and idler modes, respectively. In the case of few-mode birefringent fibers, such as the Panda-type FMF considered here (Fig. 1(a)), a specific type of FWM known as intermodal-vectorial four-wave mixing can occur. In this process, two orthogonally polarized pump photons in different spatial modes annihilate to produce two orthogonally polarized signal and idler photons, also in different spatial modes[17]. As a result, the generated signal and idler photons are correlated in the spatial-polarization-frequency DOFs[7], enabling the creation of hybrid entanglement spanning these three DOFs. To achieve this, we propose using an IV-FWM process involving four distinct fiber modes, which is known to generate two pairs of signal-idler bands[17]. In this case, the entanglement can be realized by ensuring spectral overlap between these two pairs.

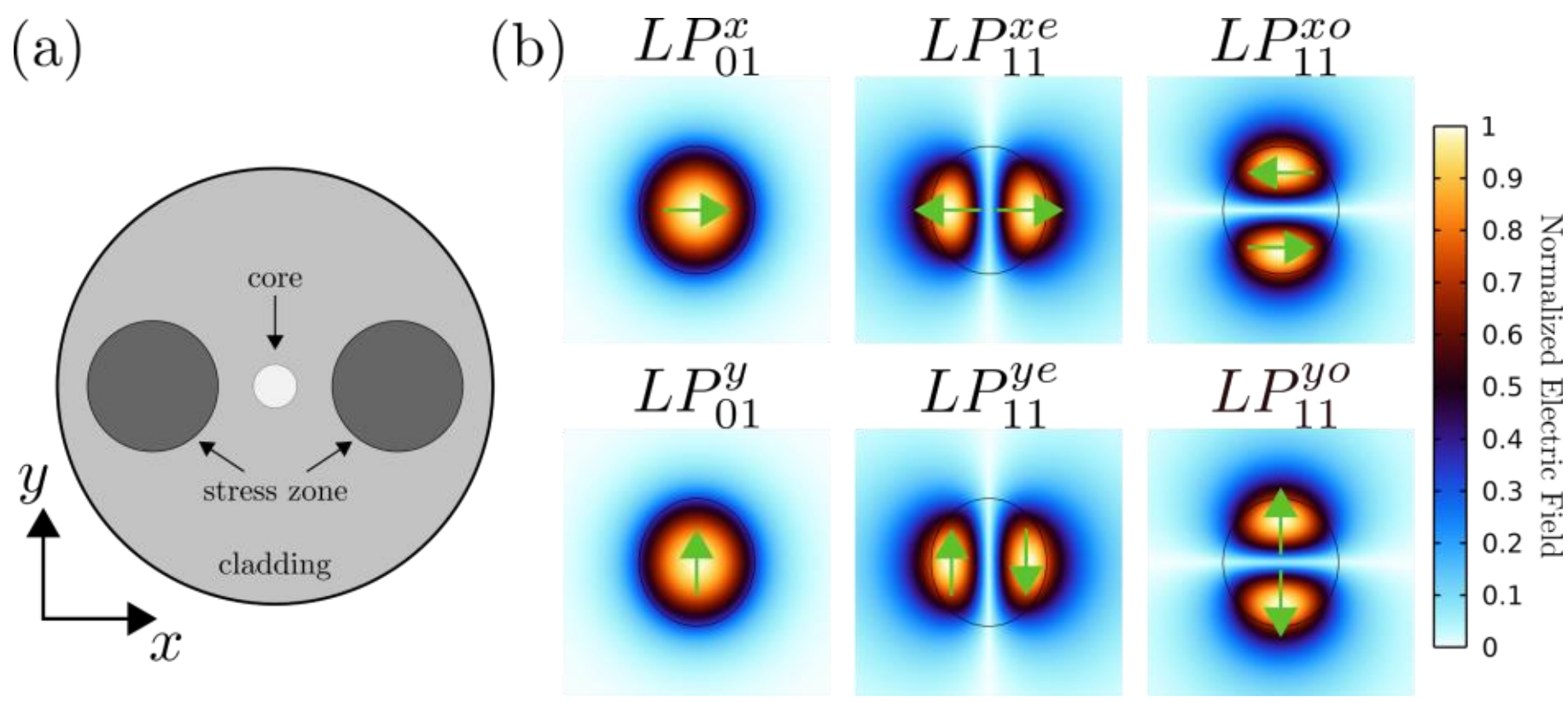

Fig. 1. (a) Schematic cross-section of the birefringent Panda-type fiber (Nufern PM1550B-XP) with stress-applying zones. (b) The absolute values of the normalized electric fields of the six guided LP modes supported by the fiber at 1064.3 nm. The green arrows indicate the direction of the electric field, i.e., the polarization of the modes.

To begin with, we examine the conditions under which the two pairs of signal-idler bands generated via IM-FWM can spectrally overlap. The occurrence of any FWM process requires phase matching among the four participating electromagnetic waves, which is mathematically expressed through their wave vectors as[18] $\beta^{(l)} + \beta^{(m)} = \beta^{(p)} + \beta^{(n)}$. For frequency-degenerate pumps, the phase-matching condition can be approximated by a second-order Taylor series expansion at the pump frequency as:

$$\frac{\beta_2^{(p)} + \beta_2^{(n)}}{2}\Omega^2 + \left(\beta_1^{(p)} - \beta_1^{(n)}\right)\Omega + \left(\beta_0^{(p)} - \beta_0^{(l)} - \beta_0^{(m)} + \beta_0^{(n)}\right) = 0, \qquad (1)$$

where $\beta_i^{(j)}$ describes the $i^{th}$ derivative of the wave vector of the $j^{th} \in \{p, l, m, n\}$ mode and the angular frequency $\Omega$ (resp. $-\Omega$) describes the spectral detuning of the signal (resp. idler) band from the pumps. Next, to build more intuition on how the linear optical properties of the fiber impact the phase matching, we rewrite Eq. 1 in the following manner:

$$-\frac{\lambda^2}{2\pi c}\underbrace{\left(\frac{D^{(p)}+D^{(n)}}{2}\right)}_{\bar{D}^{(p,n)}}\Omega^2+\frac{1}{c}\underbrace{\left(N^{(p)}-N^{(n)}\right)}_{\Delta N^{(p,n)}}\Omega+\overbrace{\left[\underbrace{\left(n^{(p)}-n^{(l)}\right)}_{\Delta n^{(p,l)}}-\underbrace{\left(n^{(m)}-n^{(n)}\right)}_{\Delta n^{(m,n)}}\right]}^{\Delta n}=0, \quad (2)$$

where $c$ is the speed of light, $D^{(p)}$ and $D^{(n)}$ (resp. $N^{(p)}$ and $N^{(n)}$) are the chromatic dispersion (resp. group refractive index) of the signal and idler modes, respectively, while $n^{(p)}, n^{(l)}, n^{(m)}, n^{(n)}$ are the phase refractive indices of the signal, $\text{pump}_1$, $\text{pump}_2$, and idler modes, respectively. Further, $\bar{D}^{(p,n)}$ and $\Delta N^{(p,n)}$ are, respectively, the average chromatic dispersion and the modal group birefringence of the signal/idler modes, while $\Delta n = \Delta n^{(p,l)} - \Delta n^{(m,n)}$ is the differential birefringence of the signal/$\text{pump}_1$ and $\text{pump}_2$/idler modes. This allows to cast the phase-matching condition in a compact form, which highlights its quadratic nature:

$$-\frac{\lambda^2}{2\pi c}\bar{D}^{(p,n)}\Omega^2+\frac{1}{c}\Delta N^{(p,n)}\Omega+\Delta n=0. \quad (3)$$

Within the framework of IV-FWM, the phase-matching condition given by Eq. (3) can be satisfied either by a two-mode or a four-mode process in a fiber supporting at least two spatial mode groups. Without loss of generality, we focus on two representative IV-FWM processes realizable by exciting specific modes of the Panda-type fiber (Fig. 1(b)): $\left(LP_{01}^{y}, LP_{11}^{xe}\right) \rightarrow \left(LP_{01}^{y}, LP_{11}^{xe}\right)$ as a two-mode process, and $\left(LP_{01}^{y}, LP_{11}^{xe}\right) \rightarrow \left(LP_{01}^{x}, LP_{11}^{ye}\right)$ as a four-mode process.

Starting with the two-mode process, where $(l,m) \rightarrow (l,m)$ and $l \neq m$, the free term $\Delta n = 0$, and Eq. 3 yields a single nonzero solution $\Omega = \frac{2\pi}{\lambda^2}\frac{\Delta N^{(l,m)}}{\bar{D}^{(l,m)}}$ for a signal band generated in the $l$ mode (here $LP_{01}^{y}$), with the corresponding $-\Omega$ for the idler band generated in the $m$ mode (here $LP_{11}^{xe}$), as shown in Fig. 2 (a). Further, when four distinct modes participate in FWM, i.e., $(l,m) \rightarrow (p,n)$, three possible scenarios can arise. First, if Eq. (3) does not yield a real solution, phase matching does not occur (Fig. 2 (b)). Second, when Eq. (3) provides real solutions and $\bar{D}^{(p,n)} \cdot \Delta n > 0$, two signal bands $(\Omega_1, \Omega_2)$ appear on one energy side of the pumps, while the corresponding idler bands $(-\Omega_1, -\Omega_2)$ appear on the opposite energy side (Fig. 2 (c)). Third, when Eq. (3) has real solutions and $\bar{D}^{(p,n)} \cdot \Delta n < 0$, one signal and one idler band are generated on each energy side of the pumps (Fig. 2 (d)). Regarding the considered $\left(LP_{01}^{y}, LP_{11}^{xe}\right) \rightarrow \left(LP_{01}^{x}, LP_{11}^{ye}\right)$ process, the signal (resp. idler) bands are generated in the $LP_{01}^{x}$ (resp. $LP_{11}^{ye}$) modes.

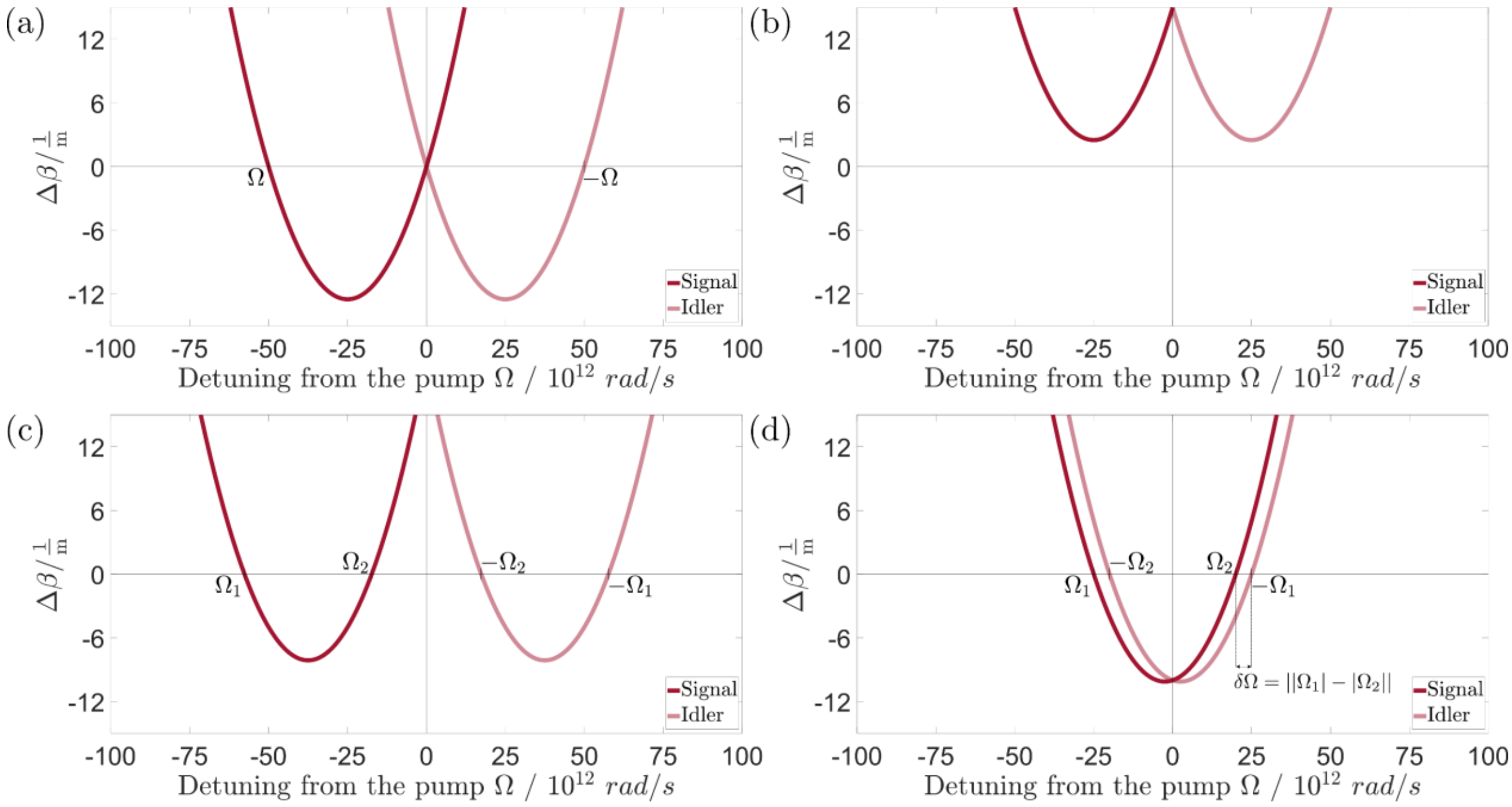

Fig. 2. Phase-matching condition of FWM calculated for four distinct cases: (a) two-mode process, (b) four-mode process, where there is no phase matching, (c) four-mode process, where both solutions $\Omega_1, \Omega_2$ appear at the same energy side of the pumps, and (d) four-mode process, where $\Omega_1, \Omega_2$ appear at the opposite energy sides of the pumps. Solid (resp. semi-transparent) lines represent the signal $\Omega$ (resp. idler $-\Omega$) solution(s). For the calculations, the following parameters were used: (a) $\bar{D} = -33.26 \frac{ps}{km \cdot nm}, \Delta N = 3e-4, \Delta n = 0$, (b) $\bar{D} = -33.26 \frac{ps}{km \cdot nm}, \Delta N = 3e-4, \Delta n = 2.54e-6$, (c) $\bar{D} = -33.26 \frac{ps}{km \cdot nm}, \Delta N = 4.5e-4, \Delta n = 3.39e-6$, and (d) $\bar{D} = -33.26 \frac{ps}{km \cdot nm}, \Delta N = 3e-5, \Delta n = -1.69e-6$.

For the third scenario shown in Fig. 2 (d), we introduce a spectral distinguishability parameter $\delta\Omega = \left||\Omega_1| - |\Omega_2|\right|$, which quantifies the frequency difference between the signal (solid line) and idler (dashed line) bands located on the same energy side of the pumps. Importantly, $\delta\Omega$ characterizes the frequency difference between the signal and idler modes, which differ in polarization and mode order. From the quadratic nature of Eq. 3, it follows directly that $\delta\Omega \to 0$ when $\Delta N^{(p,n)} \to 0$. This is confirmed in Fig. 3(a), which shows that if one reduces $\Delta N^{(p,n)}$ the signal (solid lines) and idler bands (dashed lines) converge to one another. Setting $\Delta N^{(p,n)} = 0$ (green lines), i.e., equalizing group indices of the signal and idler modes at the pump frequency, leads in the appearance of two pairs of overlapping bands located at $\Delta\Omega = \Omega_0 \pm \Omega_{1(2)}$, where $\Omega_0$ is the pump frequency. As a result, the signal and idler photons generated in the $LP_{01}^{x}$ and $LP_{11}^{ye}$ modes at $\Delta\Omega = \Omega_0 + \Omega_1$ are spectrally indistinguishable from one another and exhibit spatial-polarization correlation with the corresponding idler and signal photons created in the $LP_{11}^{ye}$ and $LP_{01}^{x}$ modes at $\Delta\Omega = \Omega_0 - \Omega_1$. This allows for the generation of two-photon states with hybrid entanglement in spatial-polarization-frequency DOFs through intermodal-vectorial FWM, under the requirement that $\Delta N^{(p,n)} \to 0$ at the pump wavelength.

From a mathematical standpoint, the spectral position of the spectrally indistinguishable bands can be tailored by keeping $\Delta N^{(p,n)} = 0$ and varying either the free term $(\Delta n)$ or the quadratic term $\left(\bar{D}^{(p,n)}\right)$ of Eq. 3, as shown in, respectively, Figs. 3 (b) and (c).

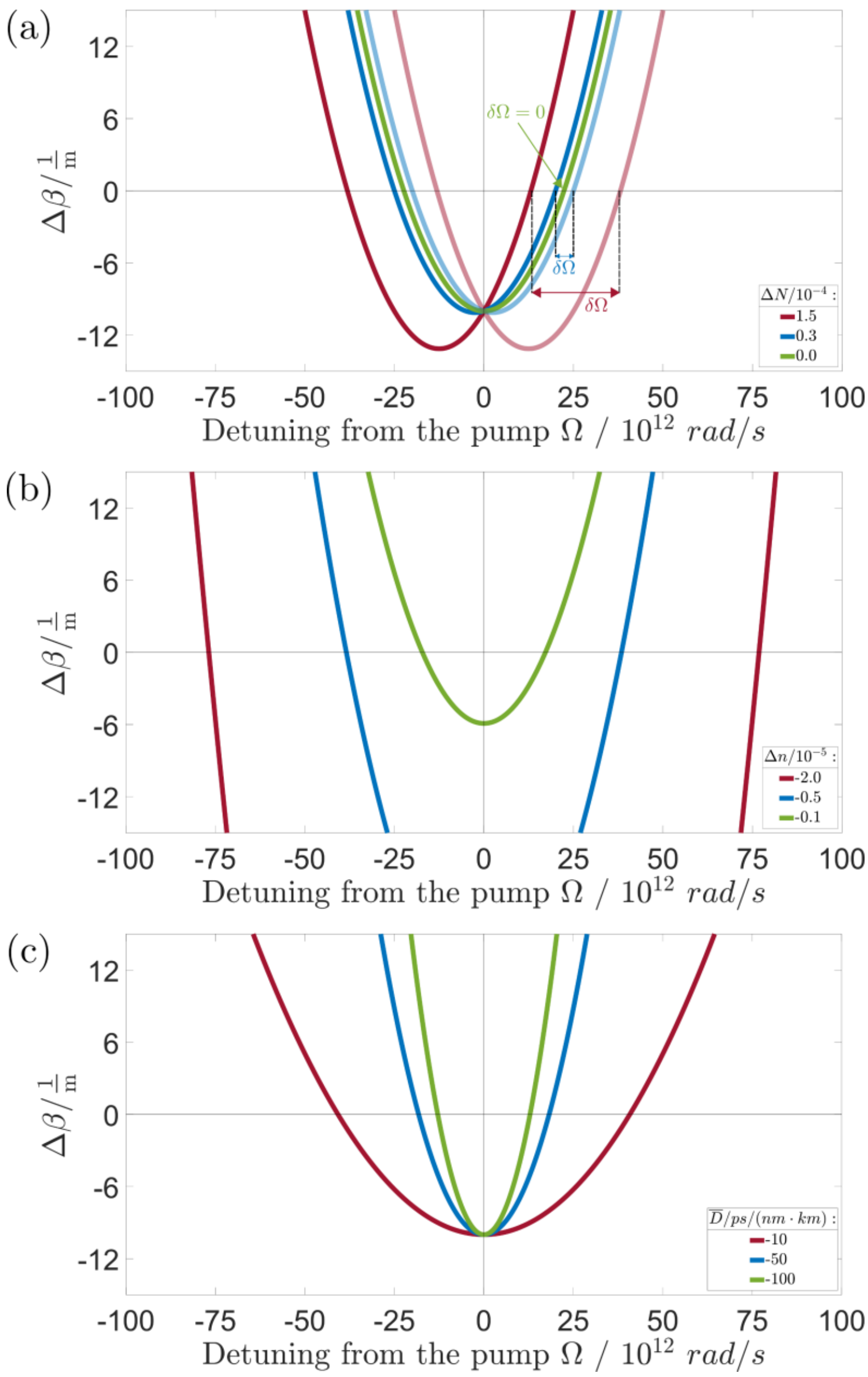

Fig. 3. Impact of the (a) modal group birefringence of the signal/idler modes $\Delta N$, (b) differential birefringence of the signal/pump$_1$ and idler/pump$_2$ modes $\Delta n$, and (c) average chromatic dispersion of the signal/idler modes $\overline{D}$ on the spectral positions of the signal-idler bands generated in a four-mode intermodal-vectorial FWM. Solid (resp. semi-transparent) lines represent the signal ($\Omega$) (resp. idler ($-\Omega$)) solution. $\delta\Omega$ is the spectral distinguishability parameter. For the calculations, the following parameters were used: (a) $\overline{D} = -33.26\frac{ps}{km\cdot nm}$, $\Delta n = 0$, (b) $\overline{D} = -33.26\frac{ps}{km\cdot nm}$, $\Delta N = 0$, and (c) $\Delta N = 0, \Delta n = -1.69e-6$.

However, from a practical point of view, it is easier to vary the differential birefringence between the modes ($\Delta n$) rather than the average chromatic dispersion $\left(\overline{D}^{(p,n)}\right)$, without significantly affecting $\Delta N^{(p,n)}$. For example, $\Delta n$ can be modified by tailoring the core ellipticity or the shape and size of stress zones inside the fiber (e.g., Panda-type, Bow-Tie). To the first approximation, such zones shift the stress-free phase refractive indices of the modes by a dispersionless constant, thereby minimally impacting the group indices hence $\Delta N^{(p,n)}$, while not impacting the spatial distribution of modes. Finally, the ability to manipulate the spectral position of such FWM is important in: (i) tailoring the process for a particular application and (ii), in terms of entanglement, moving the FWM away from the region of significant Raman scattering[18], which acts as a noise mechanisms deteriorating the signal-idler correlations[7,19].

### 3. Gain of four-wave mixing

Phase matching between a set of fiber modes is a necessary condition for FWM to occur, but it is not sufficient alone. The FWM process requires a nonzero integral overlap between the transverse components of the respective electromagnetic fields of the modes. This overlap defines the strength of the FWM process and is given by[20]:

$$S_K^{plmn} = \frac{2}{3}\left|\frac{\langle[\boldsymbol{F}_p^*\boldsymbol{F}_l][\boldsymbol{F}_m\boldsymbol{F}_n^*]\rangle}{\left[\langle|\boldsymbol{F}_p|^2\rangle\langle|\boldsymbol{F}_l|^2\rangle\langle|\boldsymbol{F}_m|^2\rangle\langle|\boldsymbol{F}_n|^2\rangle\right]^{1/2}}\right| + \frac{1}{3}\left|\frac{\langle[\boldsymbol{F}_p^*\boldsymbol{F}_n^*][\boldsymbol{F}_m\boldsymbol{F}_l]\rangle}{\left[\langle|\boldsymbol{F}_p|^2\rangle\langle|\boldsymbol{F}_l|^2\rangle\langle|\boldsymbol{F}_m|^2\rangle\langle|\boldsymbol{F}_n|^2\rangle\right]^{1/2}}\right|, \quad (4)$$

where $\boldsymbol{F}_j\,(\boldsymbol{x},\boldsymbol{y})$ represents the spatial distribution of the $j^{th} \in \{p, l, m, n\}$ fiber mode, while the integration in the brackets is carried out over the cross-section of the fiber.

To describe the relative strength of different FWM processes, it is convenient to normalize $S_K^{plmn}$ with respect to the self-overlap of the fundamental mode $S_K^{xxxx}$, i.e., $f^{plmn} = S_K^{plmn}/\,S_K^{xxxx}$. Table 1 summarizes the $f^{plmn}$ values for some representative FWM processes that can occur between the modes of the Panda-type fiber shown in Fig. 1 (b), when phase-matched. The overlaps were calculated at 1064.3 nm using COMSOL[21]. At this wavelength, the fiber supports 6 polarization modes divided into two spatial $LP_{01}$ and $LP_{11}$ groups (see Fig. 1 (b)). As shown in Table 1, the fiber supports: vectorial (V), intermodal (I), as well as two-mode and four-mode intermodal-vectorial (IV) processes, with the last one being of particular interest due to its potential spectral indistinguishability. For a detailed discussion of various FWM processes in the Panda-type fiber we refer to[17].

Table 1. Integral overlap $f^{plmn} = S_K^{plmn}/\,S_K^{xxxx}$ of selected FWM processes excitable in the fiber shown in Fig. 1.

| | $signal$ | $pump_1$ | $pump_2$ | $idler$ | |
|---|---|---|---|---|---|
| $\boldsymbol{process\ type}$ | $\boldsymbol{p}$ | $\boldsymbol{l}$ | $\boldsymbol{m}$ | $\boldsymbol{n}$ | $\boldsymbol{f^{plmn}}$ |
| $V$ | $LP_{01}^{x}$ | $LP_{01}^{x}$ | $LP_{01}^{y}$ | $LP_{01}^{y}$ | 0.667 |
| $I$ | $LP_{01}^{y}$ | $LP_{01}^{y}$ | $LP_{11}^{ye}$ | $LP_{11}^{ye}$ | 0.530 |
| $V$ | $LP_{11}^{yo}$ | $LP_{11}^{yo}$ | $LP_{11}^{xo}$ | $LP_{11}^{xo}$ | 0.529 |
| $IV$ (2 modes) | $LP_{01}^{y}$ | $LP_{01}^{y}$ | $LP_{11}^{xe}$ | $LP_{11}^{xe}$ | 0.353 |
| $IV$ (2 modes) | $LP_{11}^{xo}$ | $LP_{11}^{xo}$ | $LP_{11}^{ye}$ | $LP_{11}^{ye}$ | 0.170 |
| $IV$ (4 modes) | $LP_{11}^{ye}$ | $LP_{01}^{y}$ | $LP_{11}^{xe}$ | $LP_{01}^{x}$ | 0.353 |
| $IV$ (4 modes) | $LP_{11}^{ye}$ | $LP_{11}^{yo}$ | $LP_{11}^{xe}$ | $LP_{11}^{xo}$ | 0.167 |

Next, we establish the relationship between the strength of the FWM process $(S_K^{plmn})$ and its gain. To derive the gain, we neglect Raman scattering and assume that: (i) the pump powers are significantly higher than those of the signal and idler bands, and (ii) the pumps remain undepleted during the FWM generation. Under these assumptions, it can be shown that the gain of the FWM is given by:

$$g = \frac{-i\kappa + 4\sqrt{\left(\frac{n_2}{c}\right)^2 \omega_p \omega_n \left(S_K^{plmn}\right)^2 P_l P_m - \left(\frac{\kappa}{4}\right)^2}}{2}, \tag{5}$$

where $n_2$ is the nonlinear refractive index, $\omega_p$ (resp. $\omega_n$) is the signal (resp. idler) angular frequency, $P_l$ are $P_m$ are the pump powers, and $\kappa$ is the power-dependent effective phase-mismatch[18]. The maximum gain is achieved for $\kappa = 0$:

$$g_{max} = 2n_2 \underbrace{\sqrt{\frac{\omega_p \omega_n}{c^2}}}_{\bar{\beta}} \underbrace{\sqrt{P_l P_m}}_{\bar{P}} S_K^{plmn}, \tag{6}$$

where $\bar{\beta}$ and $\bar{P}$ represent the geometric means of the signal/idler wave vectors and the pump powers, respectively.

Let us now discuss how $\bar{\beta}$, $\bar{P}$, and $S_K^{plmn}$ impact maximum gain of FWM. Starting with $\bar{P}$ and assuming a constant total pumps power $P_0 = P_l + P_m$, it is straightforward to see that $g_{max}$ is maximized when $P_l = P_m$ and decreases with an increasingly uneven split of power between the pumps, reaching 0 when $P_{l(m)} = 0$. Second, assuming that the pumps are of the same frequency ($\omega_l = \omega_m$), $\bar{\beta}$ is maximal for $\omega_p = \omega_n = \omega_l = \omega_m$ and decreases for increasing detuning from the pumps. Nevertheless, this decrease has a negligible impact on $g_{max}$ within the $\pm 100 \times 10^{12} \frac{rad}{s}$ range around the pumps centered at 1064.3 nm, as $\bar{\beta}$ drops by only approximately 6% relative to its maximum value. Third, $g_{max}$ is a linear function of $S_K^{plmn}$ and, in our case, can vary up to four times depending on the on the modes participating in the FWM (see Table 1, $\left(LP_{11}^{yo}, LP_{11}^{xe}\right) \rightarrow \left(LP_{11}^{ye}, LP_{11}^{xo}\right)$ vs. $\left(LP_{01}^{x}, LP_{01}^{y}\right) \rightarrow \left(LP_{01}^{x}, LP_{01}^{y}\right)$ process). Therefore, under fixed excitation, the gain of FWM is primarily dependent on $S_K^{plmn}$ hence the modes selected for the process and not the frequency detuning of the signal-idler bands from the pump. Interestingly, it allows for tailoring of the spectral position of a specific IV-FWM (see Figs. 3 (b) and (c)) while keeping the gain virtually constant as a function of wavelength.

## 4. Experimental verification

In Sections 2 and 3, we defined three requirements for generating two pairs of spectrally indistinguishable signal-idler bands in four-mode intermodal-vectorial four-wave mixing. Specifically, (i) the four distinct modes participating in the process must exhibit a nonzero integral overlap (Eq. 4), (ii) the group refractive indices of the signal and idler modes must be equal at the pump frequency ($\Delta N^{(p,n)} = 0$), and (iii) the phase-matching condition (Eq. 3) must yield real solutions, which, under the criterion $\Delta N^{(p,n)} = 0$, is assured by $\bar{D}^{(p,n)} \cdot \Delta n < 0$.

To experimentally demonstrate the generation of two pairs of overlapping signal-idler bands in the Panda-type FMF (Fig. 1), we first measure the linear optical properties of the fiber to verify which IV-FWM processes meet the three aforementioned requirements. Fig. 4 (a), (b), and (c) present, respectively, the difference in phase refractive indices $\Delta n$ between modes with the same spatial distribution but different polarizations ($x$ and $y$), the group refractive indices $N$ of the modes, determined with precision up to a constant, and the chromatic dispersion $D$ of the modes supported by the fiber, experimentally measured as a function of wavelength using white-light interferometry[22,23].

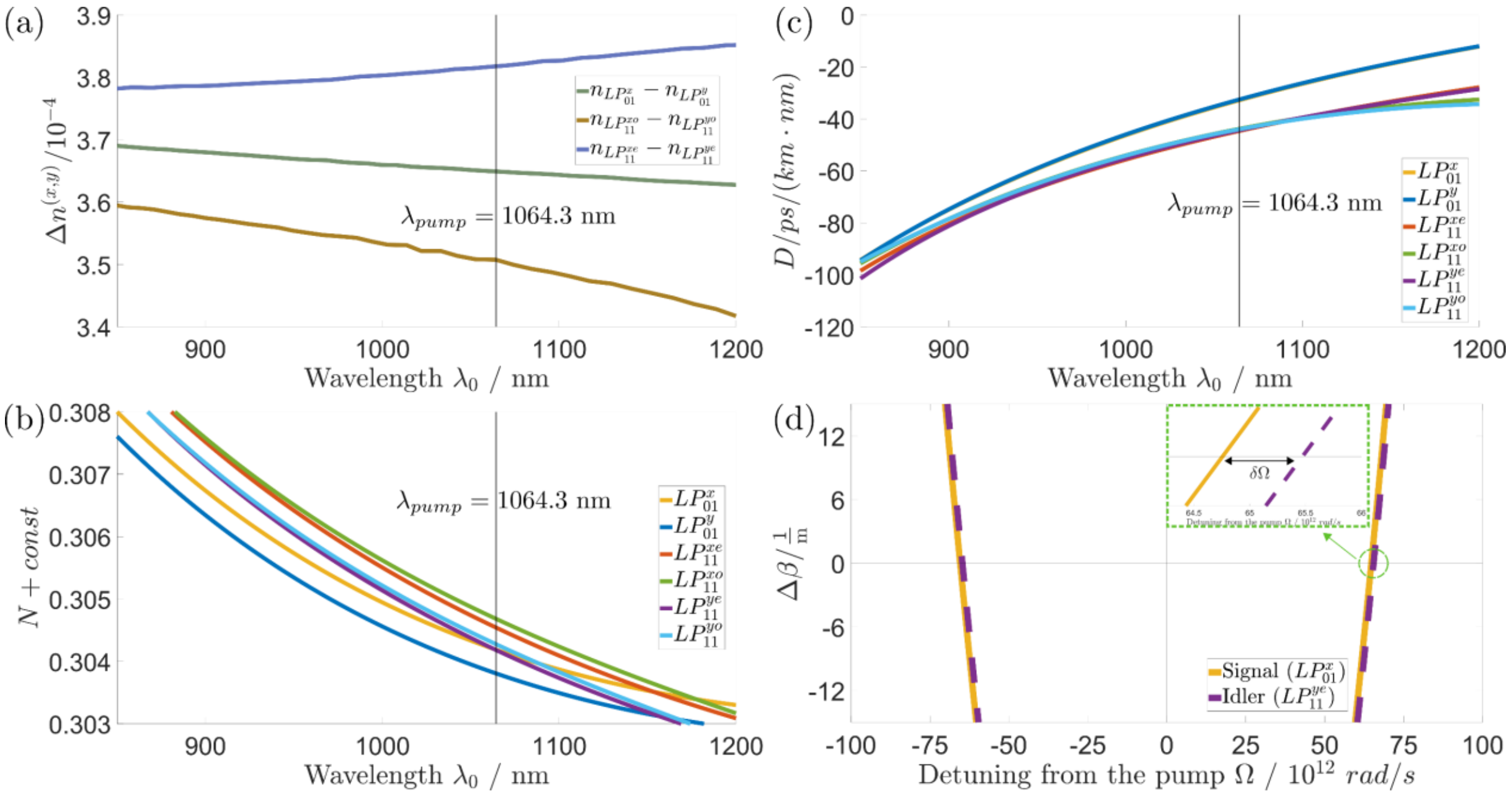

Fig. 4. Optical parameters of the Nufern 1550B-XP Panda-type fiber measured using spectral interference method[22,23]. (a) The difference in the phase refractive indices $\Delta n$ between the modes of the same spatial distribution and different polarization ($x$ and $y$). (b) The group refractive indices $N$ of the modes obtained with precision to a constant. (c) The chromatic dispersion $D$ of the modes calculated from (b). (d) The phase matching of the $(LP_{01}^{y}, LP_{11}^{xe}) \rightarrow (LP_{01}^{x}, LP_{11}^{ye})$ IV-FWM process calculated using Eq. 3 for the pumps centered at $\lambda_0 = 1064.3$ nm.

Based on these data, it can be shown that the $(LP_{01}^{y}, LP_{11}^{xe}) \rightarrow (LP_{01}^{x}, LP_{11}^{ye})$ IV-FWM process is a promising candidate for the experimental generation of two pairs of spectrally indistinguishable peaks using pumps centered at 1064.3 nm. The process has a nonzero integral overlap (Table 1), and its phase-matching (Eq. 3) results in two nearly overlapping signal-idler solutions generated in the $LP_{01}^{x}$ (purple line) and $LP_{11}^{ye}$ (yellow line) modes, as shown in Fig. 4 (d). The imperfect overlap between the solutions ($\delta\Omega \neq 0$) arises because the group refractive indices of the $LP_{01}^{x}$ (signal) and $LP_{11}^{ye}$ (idler) modes (Fig. 4 (b)) intersect at a wavelength slightly shorter than 1064.3 nm. Notably, Fig. 4 (b) shows similar crossings between the group refractive indices of different modes at various wavelengths. This creates the possibility of generating spectrally overlapping signal-idler bands through multiple four-mode IV-FWM processes within the same fiber.

To verify the above predictions, a 12 m long Panda-type FMF was pumped using an Nd:YAG laser centered at 1064.3 nm, operating at a 19 kHz repetition rate with a pulse duration of 1 ns, and an average power of 140 mW. To generate the $(LP_{01}^{y}, LP_{11}^{xe}) \rightarrow$ $\left(LP_{01}^{x}, LP_{11}^{ye}\right)$ process, the pump laser was coupled into the $LP_{01}^{y}$ and $LP_{11}^{xe}$ modes using the Wollaston prism-based setup[24].

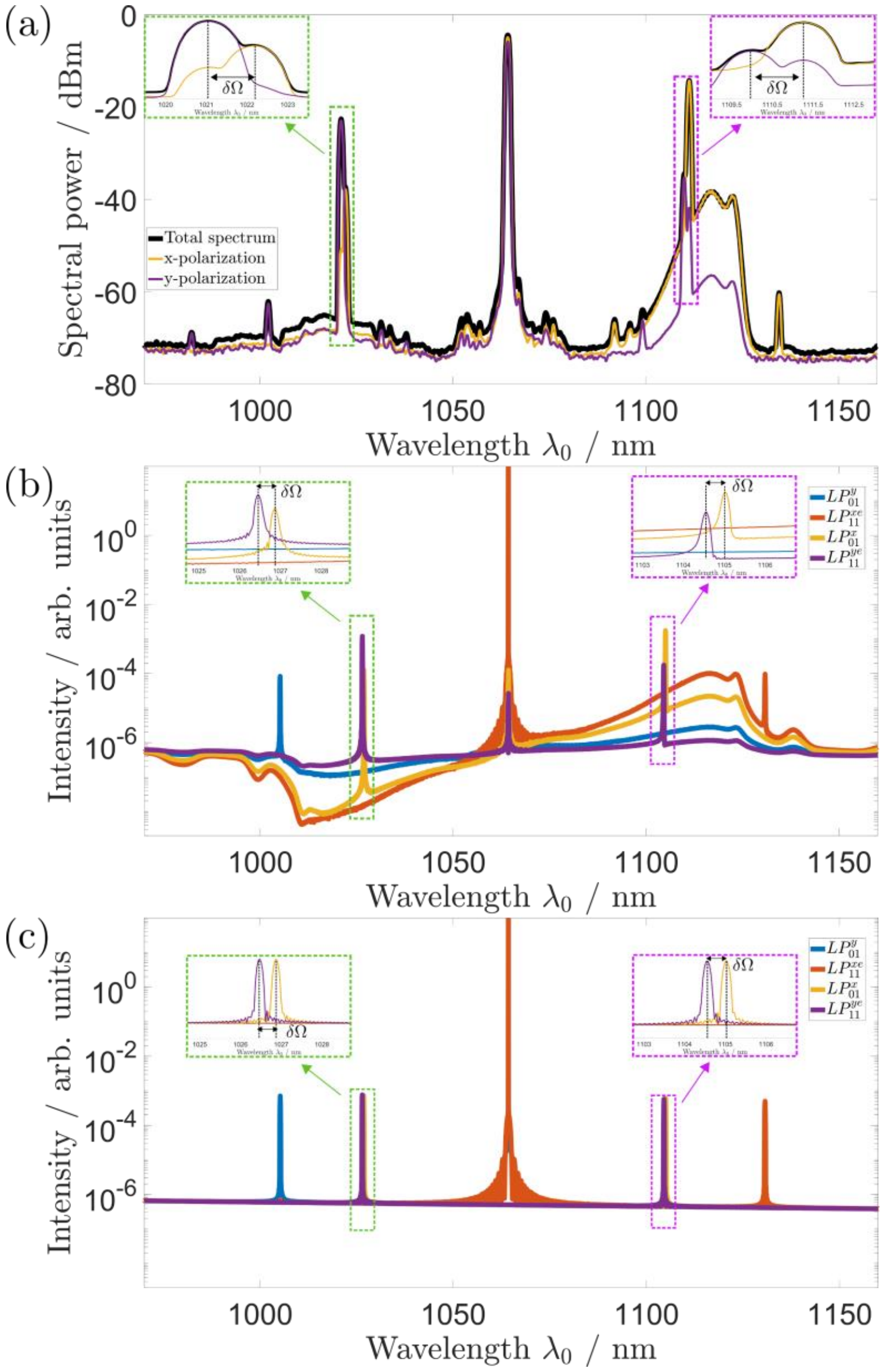

Fig. 5. (a) Polarization-resolved experimental spectra generated by pumping a Panda-type FMF with an Nd:YAG laser centered at 1064.3 nm. The pump laser was split unevenly into the $LP_{01}^{y}$ and $LP_{11}^{xe}$ modes using a Wollaston prism. (b, c) Mode-resolved numerically spectra calculated by solving the GMMNLSE, assuming a continuous-wave excitation source with a wavelength of 1064.3 nm and a power of 500 W, distributed between the $LP_{01}^{y}$ and $LP_{11}^{xe}$ modes in a 1:4 ratio. In (b) (resp. (c)) we included (resp. excluded) the Raman term in the GMMNLSE. The insets show the zoom-in on the spectrally overlapping peaks.

The measured spectra (Fig. 5 (a)) display two prominent features around 1021.5 and 1111 nm, each comprising two significantly overlapping peaks corresponding to the $x$- (yellow line) and $y$-polarized (purple line) signal components. As shown in the numerically computed mode-resolved spectra (Fig. 5 (b)), each pair of the overlapping peaks corresponds to the $LP_{01}^{x}$ and $LP_{11}^{ye}$ modes, confirming their origin in the anticipated $\left(LP_{01}^{y}, LP_{11}^{xe}\right) \rightarrow \left(LP_{01}^{x}, LP_{11}^{ye}\right)$ intermodal-vectorial four-mode FWM process. The spectral indistinguishability $\delta\Omega$ of the peaks (see the insets of Figs. 5 (a) and (b)) could be further improved ($\delta\Omega \to 0$) by using a laser with $\sim$1 nm shorter wavelength so as to match the crossing of the group refractive indices of the $LP_{01}^{x}$ and $LP_{11}^{ye}$ modes (Fig. 4 (b)). Additionally, the comparison between the experimental and numerical spectra indicates that the peaks measured at 1002 and 1134 nm are associated with the two-mode $\left(LP_{01}^{y}, LP_{11}^{xe}\right) \to \left(LP_{01}^{y}, LP_{11}^{xe}\right)$ IV-FWM. The numerical spectra were obtained by solving the Generalized Multimode Nonlinear Schrödinger Equation (GMMNLSE)[25] using the implementation provided by Chen et al.[26], extended to incorporate the Hollenbeck-Cantrell Raman model[27]. For these calculations, we used the experimentally measured linear optical properties of the fiber (Figs. 4 (a-c)) and assumed a continuous-wave excitation source with a wavelength of 1064.3 nm and a power of 500 W, distributed between the $LP_{01}^{y}$ and $LP_{11}^{xe}$ modes in a 1:4 ratio.

Further, according to Table 1 and Eq. 6, the gains of the two signal-idler bands generated in the $\left(LP_{01}^{y}, LP_{11}^{xe}\right) \to \left(LP_{01}^{x}, LP_{11}^{ye}\right)$ process, and the one generated in the $\left(LP_{01}^{y}, LP_{11}^{xe}\right) \to \left(LP_{01}^{y}, LP_{11}^{xe}\right)$ process, should be nearly identical, with only a small difference related to $\overline{\omega}$. Since the observed processes are not saturated (i.e., no cascaded FWM occurs), these equal gains should lead to equal peak heights. However, as observed in Figs. 5 (a) and (b), the two bands generated in the four-mode process not only differ in their generation rates but also exhibit significantly higher gain compared to the two-mode process. These discrepancies between the peak heights of the different bands are due to the Raman scattering (neglected in the derivation of Eq. 6), as the Raman-scattered photons are predominantly generated in the same mode as the signal responsible for the scattering[24] (see the mode-resolved Raman band in Fig. 5 (b)). Indeed, when the Raman term is excluded from the GMMNLSE (Fig. 5 (c)), the peaks of all three bands hence their gains are equal, even under uneven excitation.

Finally, it is important to emphasize that the FWMs (and their spectral positions) discussed here differ from those observed in the previous study[17], despite both studies using two Nufern PM1550B-XP fibers from the same manufacturer. This discrepancy arises from variations in the linear optical constants of the two fibers from different lots. This highlights the importance of a robust design for fibers used in nonlinear optical conversion processes like FWM.

## 5. Engineering of a hybrid entangled two-photon state

The spectral overlap of the two pairs of signal-idler bands, created simultaneously in the four-mode IV-FWM, opens up the possibility of generating photon pairs entangled in the spatial-polarization-frequency DOFs. The two-photon state associated with the realization of the $\left(LP_{01}^{y}, LP_{11}^{xe}\right) \to \left(LP_{01}^{x}, LP_{11}^{ye}\right)$ process can be written as

$$|\psi\rangle = \int d\omega_L d\omega_H \left\{ \underbrace{\eta_1 \left|LP_{11}^{ye}\right\rangle |\omega_L\rangle \left|LP_{01}^{x}\right\rangle |\omega_H\rangle}_{1^{st}\ band} + \underbrace{\eta_2 \left|LP_{01}^{x}\right\rangle |\omega_L\rangle \left|LP_{11}^{ye}\right\rangle |\omega_H\rangle}_{2^{nd}\ band} \right\}, \tag{7}$$

where $\omega_L$ (resp. $\omega_H$) corresponds to the low (resp. high) energy side of the pumps, while $\eta_1$ and $\eta_2$ are the probability amplitudes ($|\eta_1|^2 + |\eta_2|^2 = 1$) for exciting the $1^{st}$ and $2^{nd}$ pair of signal-idler bands, respectively.

When the generated bands are not saturated (as in Fig. 5), the probability amplitudes are proportional to their respective gains. This allows for the manipulation of $\eta_1$ and $\eta_2$ and consequently, the degree of entanglement of the two-photon state (Eq. 7), through the gain. Such manipulation can be achieved by leveraging Raman scattering and uneven excitation conditions, as discussed in Section 4. For example, one could design the fiber so that the generated pairs of signal-idler bands spectrally overlap with the low-gain part of the Raman band, and then vary $\eta_1$ and $\eta_2$ by adjusting the excitation ratio of the $LP_{01}^{y}$ and $LP_{11}^{xe}$ pump modes. To generate a maximally entangled Bell state, i.e., $\eta_1 = \eta_2 = \frac{1}{\sqrt{2}}$, the spectral positions of the signal-idler and Raman bands should be separated sufficiently to mitigate the impact of the latter (see Fig. 5 (c)). Since Raman scattering introduces uncorrelated noise photons[7], which are detrimental to quantum entanglement[18], whether this method can produce a state with sufficient fidelity for practical applications remains an open question.

## 6. Conclusions

To conclude, we experimentally demonstrated and theoretically explained the appearance of two pairs of spectrally overlapping signal-idler bands generated via four-mode intermodal-vectorial four-wave mixing in a birefringent few-mode fiber. We showed that spectral indistinguishability between these pairs can be achieved by pumping the fiber with a laser centered at the intersection of the group refractive indices of the signal and idler modes. The resulting two-photon state displays entanglement in the spatial-polarization-frequency degrees of freedom, and its degree of entanglement can be manipulated by spectrally tailoring the positions of the bands with respect to the Raman band and by adjusting the excitation ratio of the pump modes. During the fiber design phase, the spectral tailoring of the bands can be done by changing either the differential birefringence $\Delta n$ of the four modes participating in the FWM or the average chromatic dispersion $\overline{D}^{(p,n)}$ of the signal and idler modes. Finally, this work shows that spectrally indistinguishable intermodal-vectorial four-wave mixing generated in a few-mode birefringent fiber is a promising candidate for a source of photon pairs with hybrid entanglement in the spatial-polarization-frequency degrees of freedom.

**Funding**

Narodowe Centrum Nauki (2018/30/E/ST7/00862).